# Structural Model Updating Using Adaptive Multi-Response Gaussian Process Meta-modeling

K. Zhou

Postdoctoral Researcher

J. Tang[†]

Professor

Department of Mechanical Engineering

University of Connecticut

191 Auditorium Road, Unit 3139

Storrs, CT 06269

USA

Phone: +1 (860) 486-5911; Email: jiong.tang@uconn.edu

---

[†] Corresponding author

# Structural Model Updating Using Adaptive Multi-Response Gaussian Process Meta-modeling

K. Zhou and J. Tang[†]

Department of Mechanical Engineering

University of Connecticut

Storrs, CT 06269, USA

Phone: +1 (860) 486-5911; Email: jiong.tang@uconn.edu

**Abstract**

Finite element model updating utilizing frequency response functions as inputs is an important procedure in structural analysis, design and control. This paper presents a highly efficient framework that is built upon Gaussian process emulation to inversely identify model parameters through sampling. In particular, a multi-response Gaussian process (MRGP) meta-modeling approach is formulated that can accurately construct the error response surface, i.e., the discrepancies between the frequency response predictions and actual measurement. In order to reduce the computational cost of repeated finite element simulations, an adaptive sampling strategy is established, where the search of unknown parameters is guided by the response surface features. Meanwhile, the information of previously sampled model parameters and the corresponding errors is utilized as additional training data to refine the MRGP meta-model. Two stochastic optimization techniques, i.e., particle swarm and simulated annealing, are employed to train the MRGP meta-model for comparison. Systematic case studies are conducted to examine the accuracy and robustness of the new framework of model updating.

**Keywords:** Model updating, frequency responses, error response surface, multi-response Gaussian process (MRGP), meta-model, adaptive sampling.

## 1. Introduction

Finite element modeling is widely used in various engineering applications to facilitate structural analysis, design, and control. A reliable finite element model can simulate the load-response relations efficiently, thereby reducing the cost of experimental testing. However, discrepancies between the model and the actual structure inevitably exist, especially in terms of model parameters such as material constants, geometry and boundary conditions [1]. Hence, calibrating or updating a finite element model based on usually limited amount of experimental data from the actual structure is an important procedure.

---

[†] Corresponding author

Compared with static responses, dynamic responses are generally more sensitive to small changes of structural parameters and uncertainties, and thus commonly used for calibrating/updating investigations. Finite element model updating using dynamic/vibratory responses can be categorized as modal information based and forced response based methods. Modal information based methods employ the modal characteristics such as natural frequencies and mode shapes as inputs to update the model parameters [2-4]. It is worth noting that natural frequencies, which represent global dynamic feature, are not sensitive to local property variations. Although mode shapes may be able to capture local feature, in reality only incomplete mode shapes can be extracted through measurements, which often renders the updating difficult. On the other hand, forced response based methods, such as those using frequency response function (FRF) measurements, may overcome certain shortcomings of modal information based methods [5-7]. FRFs can be acquired at frequency bands covering resonances which are highly sensitive to structural parameter variations. They can be measured with multiple sensors in a distributed manner to capture local features especially at locations close to uncertainties. In this research we focus on finite element model updating utilizing FRFs.

Traditionally, model updating has been conducted through inverse sensitivity-based procedure which relies on an explicit relation between the change of responses and the change of structural properties/parameters. The sensitivity relations of modal information with respect to structural parameters have been derived analytically in earlier investigations [8-10]. As FRFs are commonly expressed in terms of structural parameters, e.g., stiffness, damping and mass coefficients, the sensitivity relations can also be calculated [11, 12]. Meanwhile, numerically extracting response sensitivities using finite difference provides an alternative approach [13]. With the sensitivity relation, an inverse problem can then be solved utilizing the difference between model prediction and response measurement under the same operating condition. Although relatively straightforward, the accuracy of these inverse sensitivity-methods is influenced by a number of factors. First, the number of unknown model parameters to be updated and the number of measured responses usually differ, which leads to either overdetermined or underdetermined problems. The sensitivity matrix may be ill-conditioned. Second, when only the linear term in the sensitivity relation is retained, the effectiveness of these methods is limited to the close vicinity of the nominal values of the parameters to be updated, the range of which may not be known *a priori*. Retaining higher-order terms will increase significantly the numerical cost of the inverse analysis. Third, measurement noise and uncertainties further compound the aforementioned issues. Since FRF responses are sensitive to parametric variation, the inverse analysis presents significant challenges.

A different and new school of thoughts is to formulate an optimization based inverse identification, which aims at minimizing the difference between model prediction (under sampled unknown parameters) and the measurement. In the past, a brute force Monte Carlo type repeated simulations of the finite element model in the unknown parametric space would yield prohibitive computational cost. However, the recently dramatic advancement in statistical inference has provided viable tools to produce meta-models employing a much reduced number of simulations, which shows extremely positive prospect. A meta-model, once established, can rapidly predict responses/variations of a model upon given parameters. This leads to an efficient way of evaluating the objective function or response surface in the model updating problem as unknown model parameters vary. In recent years, several parametric and non-parametric meta-models have been explored in structural dynamic analysis toward the mission of finite element model updating. Wan and Ren [14] developed a Gaussian process (GP) model to capture the relation between response residual and model parameters, which facilitated the parameter selection in terms of global analysis of GP response surface. Machado et al [15] employed a spectral approach, i.e., Karhunen-Loeve expansion to conduct model updating for damage quantification under uncertainties. Park et al [16] adopted a neural network approach to update the boundary condition of a finite element model. Zhang and Hou [17] proposed a support vector machine (SVM)-based response surface method to update parameters through minimizing the reserved singular values.

A major question in meta-model based inverse identification is the fidelity of the response surface produced and the associated optimization procedure. In order to find the true combination of unknown parameters, the response surface has to be accurate. While oftentimes meta-model accuracy is validated by using a small subset of holdout testing samples in many other applications, it is practically difficult to validate the response surface in the entire parametric space in the problem of model updating, especially when the output/error is sensitive to parametric variation. In some cases, the unknown model parameters to be updated are high-dimensional, which poses further challenge in properly carrying out the sampling based optimization. To tackle this issue, a series of efforts have been made. Sun et al [18] developed a hierarchical Bayesian framework with Laplace priors to update the finite element model using response functions extracted from ambient noise measurements. Zhou and Tang [19] formulated a Bayesian-inference stochastic model updating framework, in which Markov Chain Monte Carlo (MCMC) is adopted to expedite the parameter optimization based upon the Bayesian posterior probability density function (PDF). Shabbir and Omenzetter [20] proposed the combination of genetic algorithm and sequential niche technique to conduct parametric optimization in finite element model updating. In these approaches, guided sampling is practiced. Indeed, some stochastic optimization algorithms inherently utilize the fundamental thought of MCMC [21]. Nevertheless, many of these aforementioned optimization processes consider single output response and allow only one parameter to be sampled at each iteration. This overlooks the natural correlation between different output variables in finite element prediction, and may result in slow convergence and being trapped to local extrema in model updating.

In this research, we aim at establishing an efficient and accurate computational framework for finite element model updating using FRF measurement. We leverage upon the power of meta-model, and utilize it to develop response surface which is the error between meta-model prediction and actual measurement. A combination of inter-related improvements will be devised. We adopt the recent advancement in meta-modeling, the multi-response Gaussian process (MRGP) strategy [22, 23], so we can effectively predict the frequency responses at multiple locations that take into consideration of their inherent correlation. We develop an adaptive sampling strategy to reach high accuracy of MRGP meta-model with reduced computational burden in repeated finite element simulations. Specifically, we start from a small size of randomly selected samples to establish the initial response surface. Subsequently, we identify regions with small errors between model predictions and measurements, and generate new parametric samples in the corresponding parametric subspaces to carry out further finite element simulations. The newly acquired simulation results will be used to enrich the training dataset for the re-training of MRGP meta-model. This process continues until the optimal set of unknown parameters is identified. Meanwhile, we incorporate two representative stochastic optimization procedures, particle swarm [24] and simulated annealing [25], into MRGP training. The aforementioned adaptive sampling strategy can be conveniently integrated together with these procedures, and the resulting performances are compared.

The rest of the paper is organized as follows. In Section 2, the proposed computational framework of finite element model updating utilizing FRF is outlined, where the finite element model, its relation with respect to frequency response, the MRGP formulation, and the computational procedure for inverse identification are explained in detail. Section 3 illustrates the actual implementation through case studies on a benchmark plate structure. The role of adaptive sampling and stochastic optimization are demonstrated systematically. Section 4 provides concluding remarks.

## 2. Model Updating Framework through Meta-modeling

In this section, we present the model updating framework. We assume that a) the baseline finite element model with parameters to be updated has been established; and b) frequency response measurements at multiple locations are available. Our strategy is to formulate MRGP meta-model utilizing a limited amount of finite element simulations as training dataset to develop a response surface that characterizes the discrepancy between model prediction under sampled parameters and the actual measurement, and identify model parameters using stochastic optimization.

### *2.1. Finite element model with parameters to be updated and frequency response based formulation*

We consider a linear vibration system modeled using finite element discretization with $N$ degrees-of-freedom (DOFs). Initially, we have the *baseline* model,

$$\mathbf{M}\ddot{\mathbf{z}} + \mathbf{C}\dot{\mathbf{z}} + \mathbf{K}\mathbf{z} = \mathbf{f} \tag{1}$$

where $\mathbf{z}$ is the $N$-dimensional displacement vector, $\mathbf{M}$, $\mathbf{C}$, and $\mathbf{K}$ are, respectively, the mass, damping, and stiffness matrices of dimension ($N \times N$), and $\mathbf{f}$ is the time-dependent, $N$-dimensional external excitation force vector. We assume that the structure is lightly damped with proportional damping, and thus we focus on updating the mass and stiffness matrices only. Without loss of generality, we divide the finite element model of the entire structure into $m$ segments, and let $\mathbf{K}_i$ and $\mathbf{M}_i$ denote respectively the stiffness and mass matrices of the $i$-th segment of the baseline model. Each segment corresponds to one parameter to be updated for the stiffness matrix and one parameter to be updated for the mass matrix. The stiffness and mass matrices of the *actual* structure can thus be expressed as [8]

$$\hat{\mathbf{K}} = \sum_{i=1}^{m} \mathbf{K}_i (1 + \alpha_i) \tag{2a}$$

$$\hat{\mathbf{M}} = \sum_{i=1}^{m} \mathbf{M}_i (1 + \gamma_i) \tag{2b}$$

where $\alpha_i$ and $\gamma_i$ represent respectively the stiffness and mass variation coefficients of the $i$-th segment to be identified/updated based on frequency response measurement. As the stiffness and mass matrices are generally positive definite, $\alpha_i$ and $\gamma_i$ fall into $[-1, \infty]$. We let $\boldsymbol{\alpha} = [\alpha_1, \alpha_2, \cdots, \alpha_m]$ and $\boldsymbol{\gamma} = [\gamma_1, \gamma_2, \cdots, \gamma_m]$. The equation of motion of the actual e structure with parameters to be updated can be expressed as

$$\hat{\mathbf{M}}(\boldsymbol{\gamma})\ddot{\mathbf{z}} + \hat{\mathbf{C}}(\boldsymbol{\alpha}, \boldsymbol{\gamma})\dot{\mathbf{z}} + \mathbf{K}(\boldsymbol{\alpha})\mathbf{z} = \mathbf{f} \tag{3}$$

In this research, we use frequency responses measured from actual structure to update the above finite element model. Frequency responses can be acquired at multiple locations. They are sensitive to parametric variations especially near structural resonances. Let us consider a harmonic excitation $\mathbf{f}(t) = \mathbf{F}e^{j\omega t}$ where $\mathbf{F}$ is a constant vector of force magnitude and $\omega$ is the sweeping frequency. We then have the vector-form frequency response function of the structure as

$$\mathbf{Z} = [-\omega^2 \hat{\mathbf{M}}(\boldsymbol{\gamma}) + j\omega \hat{\mathbf{C}}(\boldsymbol{\alpha}, \boldsymbol{\gamma}) + \mathbf{K}(\boldsymbol{\alpha})]^{-1} \mathbf{F} \tag{4}$$

where $\mathbf{Z}$ is the vector-form response amplitude of the entire structure.

The frequency response predicted by the finite element model under sampled unknown parameters will be compared with actual measurement to facilitate model updating. In actual practice, only a small number of sensors are employed [19]. We thus assume only a subset of the total DOFs can be measured. Correspondingly, $\mathbf{u}$, a subset of $\mathbf{Z}$, will be used in model updating, and its dimension is $n$ ($n \ll N$). We further assume that frequency responses are measured at $p$ discrete frequency points, $\boldsymbol{\omega} = [\omega_1, \omega_2, \cdots, \omega_p]$, as the measurements are collected during frequency sweep. The differences between model prediction and measurement at different frequencies can then be collectively written in the following vector form,

$$\Delta \mathbf{u}(\boldsymbol{\alpha}, \boldsymbol{\gamma}) = [\Delta u_{1,1}, \cdots \Delta u_{n,1}, \Delta u_{1,2}, \cdots, \Delta u_{n,2}, \cdots\cdots, \Delta u_{1,p}, \cdots, \Delta u_{n,p}]^T \tag{5}$$

Our goal is to find ($\boldsymbol{\alpha}, \boldsymbol{\gamma}$) such that the difference between model prediction and measurement is minimized. Indeed, there is a variety of metrics that can be used to define the closeness of the model prediction under sampled parameters and the measurement. This may lead to different mathematical formulations of optimization. For example, we may adopt different norms of $\Delta \mathbf{u}(\boldsymbol{\alpha}, \boldsymbol{\gamma})$, and even set up a multi-objective optimization problem where individual differences listed in Equation (5) are all treated as separate objectives. Owing to the similar nature of frequency response differences at different locations and in order to focus on the establishment of the new computational framework, in this research we formulate a single objective optimization in the following manner to identify the unknown parameters,

$$\text{Find: } \boldsymbol{\alpha}^* = [\alpha_1^*, \alpha_2^*, \cdots, \alpha_m^*], \quad \boldsymbol{\gamma}^* = [\gamma_1^*, \gamma_2^*, \cdots, \gamma_m^*], \quad \alpha_i^*, \beta_i^* \in [-1, \infty] \tag{6a}$$

$$\text{Minimize } \varepsilon_e, \text{ where } \varepsilon_e = \frac{1}{n \cdot p} \left( \sum_{i=1}^{n} \sum_{j=1}^{p} \left| \frac{\Delta u_{i,j}}{\bar{u}_{i,j}} \right| \right) \tag{6b}$$

where $\bar{u}_{i,j}$ is the response amplitude measured at the *i*-th DOF of the actual structure under the *j*-th excitation frequency. Each item in the summation represents the respective error percentage. We plan to establish error response surface with respect to unknown parameters to facilitate model updating.

## *2.2. Error response surface construction using multi-response Gaussian process (MRGP)*

Since each finite element simulation under sampled unknown parameters is computationally costly and we oftentimes do not have *a priori* knowledge of how sensitive the system response is with respect to specific parametric combinations, the brute force Monte Carlo approach cannot be applied here to construct the error response surface. Here we resort to Gaussian process (GP) which typically uses a much smaller sample size of training dataset. In particular, we adopt the multi-response Gaussian process (MRGP) strategy that is capable of emulating frequency responses at multiple sensor locations. This avoids the burden of training multiple single-response Gaussian processes, and further allows us to capture inherent correlation of frequency responses at different locations.

In Gaussian process formulation, a system is generally denoted as $g(\mathbf{x})$, in which **x** is an input vector. In this research, **x** represents the sampled model parameters. The observed output is **y** is also a vector, and here it represents the multiple frequency responses. In other words, the input $\mathbf{x}_i$ and output $\mathbf{y}_i$ represent the *i*-th sampled parameter vector $[\boldsymbol{\alpha}, \boldsymbol{\gamma}]$ (Equation (2)) and the corresponding response error vector $\Delta \mathbf{u}$ (Equation (5)), respectively. Given a set of *r* sampled observations, considered as training dataset $(\mathbf{X}, \mathbf{Y}) = \{(\mathbf{x}_i, \mathbf{y}_i), i = 1, 2, \ldots r\}$, as described below, a multi-response Gaussian process (MRGP) regression will be established that can be further employed to predict the output $\mathbf{Y}^*$ over target input $\mathbf{X}^*$ where $\mathbf{X}^*$ and $\mathbf{Y}^*$ are matrices.

The prior of MRGP is expressed as

$$\mathbf{Y} \sim \mathrm{GP}(\mathbf{H}(\mathbf{X})\boldsymbol{\beta}, \mathbf{Q}\boldsymbol{\Sigma}(\mathbf{X}, \mathbf{X}')) \tag{7}$$

The dimensions of the input vector $\mathbf{x}$ and the output vector $\mathbf{y}$ are, respectively, $2m$ and $n \times p$. $\mathbf{H}(\mathbf{X})\boldsymbol{\beta}$ denotes the mean function of MRGP of observed data. In this study, without loss of generality the linear mean function is adopted, which yields $\mathbf{H}(\mathbf{X})$ in the form of $\mathbf{H}(\mathbf{X}) = \begin{bmatrix} 1 & x_{1,1} & \ldots & x_{1,2m} \\ 1 & x_{2,1} & \ldots & x_{2,2m} \\ \ldots & \ldots & \ldots & \ldots \\ 1 & x_{k,1} & \ldots & x_{k,2m} \end{bmatrix}$, where $x_{i,j}$ is $j$-th element of the $i$-th sample of the input matrix $\mathbf{X}$. $\boldsymbol{\beta}$ is the unknown regression matrix, in which $\boldsymbol{\beta} = \begin{bmatrix} \beta_{1,1} & \beta_{1,2} & \ldots & \beta_{1,n\times p} \\ \beta_{2,1} & \ldots & \ldots & \beta_{2,n\times p} \\ \ldots & \ldots & \ldots & \ldots \\ \beta_{2m,1} & \ldots & \ldots & \beta_{2m,n\times p} \end{bmatrix}$. $\mathbf{Q}$ is the non-spatial correlation matrix to account for the statistical correlation of frequency responses acquired, and $\boldsymbol{\Sigma}$ is a spatial covariance matrix that is determined by the specified covariance function. The dimensions of $\mathbf{Q}$ and $\boldsymbol{\Sigma}$ are, respectively, $np \times np$ and $r \times r$. Here we select the isotropic squared exponential covariance function, which is expressed as

$$k(\mathbf{x}_i, \mathbf{x}_j) = \sigma_f^2 e^{-\frac{(\mathbf{x}_i - \mathbf{x}_j)^T (\mathbf{x}_i - \mathbf{x}_j)}{2l}} + \sigma_n^2 \delta_{ij} \tag{8}$$

where $k(\mathbf{x}_i, \mathbf{x}_j)$ is the resulted value that represents the entry at the $i$-th row and the $j$-th column of covariance matrix $\boldsymbol{\Sigma}$. In this study, $\sigma_n$ is set as zero for simple case without noise. As will become clear later, $\boldsymbol{\beta}$ is dependent on $\sigma_f^2$ and $l$. We thus refer to $\phi = [\sigma_f^2, l]$ as the hyper-parameters of MRGP meta-model, which are to be optimized using training dataset. Signal variance $\sigma_f^2$ is a scaling factor, representing the variation of function values from their mean. Small values of $\sigma_f^2$ indicate functions that stay close to their mean value, and larger values allow more variation. Lengthscale $l$ describes how smooth a function is. Small lengthscale values mean that function values can change quickly, whereas large values characterize functions that change slowly. These hyper-parameters can be obtained from maximum likelihood estimation (MLE) [28] in terms of multivariate normal distribution shown below,

$$p(\mathbf{Y}|\mathbf{X},\phi) = (2\pi)^{-\frac{r.n.p}{2}}(\det\mathbf{Q})^{-\frac{n.p}{2}}(\det\mathbf{\Sigma})^{-\frac{r}{2}}\exp\left\{-\frac{1}{2}\text{vec}(\mathbf{Y}-\mathbf{H}\boldsymbol{\beta})^T(\mathbf{Q}\otimes\mathbf{\Sigma})^{-1}\text{vec}(\mathbf{Y}-\mathbf{H}\boldsymbol{\beta})\right\} \quad (9)$$

where vec(.) denotes the vectorization operator to convert a 2-dimensional matrix to 1-dimensional vector, and $\otimes$ denotes the Kronecker product of two matrices. The conventional way to maximize the above likelihood is to maximize the corresponding logarithm likelihood given as

$$\ln(p(\mathbf{Y}|\mathbf{X},\phi)) = -\frac{r.n.p}{2}\ln(2\pi) - \frac{r}{2}\ln(\det\mathbf{Q}) - \frac{n.p}{2}\ln(\det\mathbf{\Sigma}) - \frac{1}{2}\text{vec}(\mathbf{Y}-\mathbf{H}\boldsymbol{\beta})^T(\mathbf{Q}\otimes\mathbf{\Sigma})^{-1}\text{vec}(\mathbf{Y}-\mathbf{H}\boldsymbol{\beta}) \quad (10)$$

We can obtain the optimized $\hat{\boldsymbol{\beta}}$ by setting the derivative of the above expression with respect to $\boldsymbol{\beta}$ to be zero. $\hat{\boldsymbol{\beta}}$ is calculated as

$$\hat{\boldsymbol{\beta}} = [\mathbf{H}^T\mathbf{\Sigma}(\phi)\mathbf{H}]^{-1}\mathbf{H}^T\mathbf{\Sigma}(\phi)\mathbf{Y} \quad (11)$$

As $\hat{\boldsymbol{\beta}}$ is the function of variable $\mathbf{\Sigma}(\phi)$, the hyper-parameters $\phi = [\sigma_f^2, l]$ are the only ones that affect the model training. The optimized correlation matrix is then derived as [22]

$$\hat{\mathbf{Q}} = \frac{1}{r}(\mathbf{Y}-\mathbf{H}\hat{\boldsymbol{\beta}})^T\mathbf{\Sigma}(\phi)^{-1}(\mathbf{Y}-\mathbf{H}\hat{\boldsymbol{\beta}}) \quad (12)$$

It can be seen that $\hat{\mathbf{Q}}$ is also dependent on the hyper-parameters $\phi = [\sigma_f, l]$.

After we substitute Equations (11) and (12) into Equation (10), the best hyper-parameters $\hat{\phi} = [\hat{\sigma}_f^2, \hat{l}]$ can be identified, which in turn yields optimized $\hat{\mathbf{\Sigma}}$. A variety of stochastic optimization techniques have been attempted to find the hyper-parameters in Gaussian process meta-modeling [29]. In the subsequent section we will compare the performances of two representative algorithms for the specific application of error surface construction in finite element model updating. Once the MRGP meta-model is trained, the predicted output $\mathbf{Y}^*$ of the target input $\mathbf{X}^*$ based upon the posterior of MRGP is given as

$$\mathbf{Y}^*|\mathbf{Y},\mathbf{X},\mathbf{X}^*,\hat{\phi} \sim \text{GP}(\boldsymbol{\mu},\boldsymbol{\Omega}) \quad (13)$$

where the posterior mean function and covariance under the optimized hyper-parameters are, respectively,

$$\boldsymbol{\mu} = \mathbf{H}(\mathbf{X}^*)\hat{\boldsymbol{\beta}} + \hat{\mathbf{\Sigma}}^{*T}\mathbf{\Sigma}^{-1}(\mathbf{Y}-\mathbf{H}\boldsymbol{\beta}) \qquad \boldsymbol{\Omega} = \hat{\mathbf{Q}}\otimes(\hat{\mathbf{\Sigma}}^{**} - \hat{\mathbf{\Sigma}}^{*T}\hat{\mathbf{\Sigma}}^{-1}\hat{\mathbf{\Sigma}}^*) \quad (14)$$

where $\hat{\mathbf{\Sigma}}^* = \hat{\mathbf{\Sigma}}(\mathbf{X},\mathbf{X}^*)$ and $\hat{\mathbf{\Sigma}}^{**} = \hat{\mathbf{\Sigma}}(\mathbf{X}^*,\mathbf{X}^*)$. It's worth noting that the MRGP training efficiency depends on the number of training data and the dimension of responses, which determine the sizes of non-sparse matrices, i.e., $\mathbf{\Sigma}$ and $\mathbf{Q}$, respectively.

### 2.3. Computational procedure with adaptive sampling and meta-model refinement

The foundational idea of the proposed framework is to establish the error response surface under sampled unknown model parameters, i.e., the objective function, by means of MRGP. This error response surface is then utilized to identify the model parameters that yields the minimal error with respect to frequency response measurement. The key is to establish efficiently and accurately the error response surface that can capture all the minima with sufficient resolution. For this purpose, we develop a procedure with adaptive and guided sampling, and meta-model refinement as explained below. We assume we have the baseline finite element model and the set of model parameters to be updated. We also assume that we have acquired the frequency response measurements from the actual structure. It involves the following three steps.

1. We first build an initial MRGP meta-model. We sample $q$ sets of model parameters. Assuming we only know the ranges of these parameters (which are to be identified/updated), we use uniform distribution. Subsequently, we carry out finite element simulations (Equation (4)) to obtain the frequency responses under the sampled parameters. For each simulation, we compare the frequency response with the measurement to obtain the corresponding error $\varepsilon_e$ (Equation (6b)). In the unlikely event that the error level is already below a pre-specified threshold $\varepsilon_s$, the model parameters will have been identified and we terminate the process. Otherwise, these $q$ sets of input-output relations (i.e., sampled model parameters and the subsequent frequency responses at pre-specified DOFs and frequency points) are used as training data to form the MRGP meta-model, following the algorithm outlined in Section 2.2. This meta-model can be readily used to construct the initial error response surface in the parametric space through efficient emulation.

2. We proceed to an iterative process. We use the MRGP meta-model trained to predict frequency response in the parametric space. Since MRGP meta-model is highly efficient, we can quickly predict frequency responses under $w$ sets of parameters throughout the parametric space, and then construct the current-stage error response surface by comparing MRGP prediction results with the actual frequency response measurement. We then identify parametric regions exhibiting smaller error values. Within these regions with smaller error values, we sample an additional $s$ sets of parameters (to refine the search). In order for the samples produced to be able to cover different local minima, we impose a sampling constraint that the minimal distance of any two samples at the same iteration must exceed a certain distance threshold. We carry out further finite element simulations upon these additional $s$ sets of sampled parameters and obtain corresponding errors. If an error is below the threshold $\varepsilon_s$, we find the model parameters to be updated. Otherwise, we then re-train the MRGP meta-model, using all the input-output relations we have accumulated (i.e., the $q$ data used in the preceding MRGP training and the additional $s$ data generated in this current stage). This improved meta-model can be readily used to construct the refined error response surface in the parametric space.

3. Care should be taken as we continue using the improved MRGP meta-model for parametric identification. As indicated in Equation (6a), initially the range of parameters to be updated may fall into $[-1, \infty]$. As we proceed throughout this process, we need to narrow down the ranges of the parameters involved in model updating. To accomplish this, we sort all the accumulated ($q+s$) sets of parameters in terms of errors provided by the corresponding finite element simulations, and find the best $z$ samples with the smallest errors. We calculate the mean and variance of these $z$ sampled inputs (model parameters), based upon which a normal distribution (with updated means and variances) of these model parameters is established. Using the normal distribution of model parameters, we then generate $w$ sets of parameters within the parametric space that has been narrowed down. We go to Step 2 mentioned above and repeat the computational procedure.

Figure 1 shows the flowchart of the procedure. It features two guided sampling steps that are inter-related: 1) enriching finite element simulations within the regions that exhibit smaller errors of MRGP emulation with respect to actual measurement; and 2) narrowing down the parametric space for targeted search of unknown model parameters. Table 2 lists the implementation steps with actual computational parameters employed in a case demonstration, which will be further discussed in the following section.

## 3. Methodology Demonstration and Case Analysis

In this section we demonstrate the new framework through case investigations. We illustrate the MRGP meta-modeling and the subsequent model updating. In order to provide insights to implementation details, we compare two stochastic optimization approaches that are used to facilitate the framework, and examine the accuracy and robustness of the new method.

### 3.1. Model setup of the case investigation

A benchmark multi-plate structure, shown in Figure 2, is analyzed. The material constants of the baseline structure without variation/uncertainties are: Young's modulus $2.06 \times 10^{11}$ Pa, mass density 7850 Kg/m$^3$ and Poisson's ratio 0.3. Proportional damping is assumed, $\mathbf{C} = a\mathbf{M} + b\mathbf{K}$, and here $a = 10^{-2}$ and $b = 10^{-4}$. We use 8-node solid element in discretization. The finite element model has 3,510 DOFs. We choose this structural configuration so interested readers can readily re-construct the mesh for validation and comparison. As can be observed, this benchmark structure consists of three smaller plates joined together, which resembles topologies of complex engineering structures consisting of multiple substructures. For illustration, we divide the structure into 6 segments, each containing one stiffness parameter to be updated (Equation (2)). For the sake of demonstration and validation, the actual stiffness variations of 6 segments are given as $\bar{\boldsymbol{\gamma}} = [-0.6, -0.1, -0.3, -0.2, -0.2, -0.4]$ in this case study. Here, we purposely formulate a problem with large stiffness variations to be identified. Such kind of problem may not be solved effectively by sensitivity-based methods. The mass parameters are assumed to be accurate and not subjected to updating. Hence the number of model parameters to be identified/updated is 6. The response measurement is numerically generated utilizing the finite element model with actual stiffness variation values. In this case demonstration we resort to numerical simulation which is capable of providing the true values of the underlying parameter variations.

As shown in the Figure 2, we place 6 sensors at designated locations to acquire the *z*-direction frequency responses. Therefore, the frequency responses at a total of 6 DOFs are measured. Without loss of generality, harmonic forces with unit amplitude are applied at these locations, and the excitation frequency goes through sweep to generate frequency responses. In the subsequent case demonstration, we use numerically produced frequency responses as information employed for model updating. The first two natural frequencies of *z*-direction bending modes of the actual structure are 112.12 Hz and 291.35 Hz, respectively. As frequency response functions are generally more sensitive to parametric variations around the resonant frequencies, we pick a total of 12 frequency points, i.e., 106 Hz, 108 Hz, 110 Hz, 112 Hz, 114 Hz, and 116 Hz around the first natural frequency, and 285 Hz, 287 Hz, 289Hz, 291Hz, 293 Hz, and 295 Hz around the second natural frequency, to acquire the corresponding frequency responses. Thus, in Equation (5), $n=6$ and $p=12$. While here we resort to numerical simulation for methodology validation, in actual experimental implementation we can first identify the natural frequencies of the structure, followed by acquiring frequency responses at a selected number of frequency points. These can be realized by frequency sweep. In this research, we use finite element code developed by ourselves using MATLAB to carry out the investigation. This will facilitate a streamlined process for the adaptive training of MRGP meta-model.

*3.2. Role of stochastic optimization*

MRGP meta-modeling is the foundation of this new model updating framework. To establish the MRGP meta-model through the proposed adaptive and guided sampling strategy, care should be taken in the training process to identify hyper-parameters. As can be seen in Equation (10), the evaluation of objective function requires the calculation of the inverses and determinants of covariance matrices $\Sigma$ and $\mathbf{Q}$. These matrices in some cases are extremely singular or ill-conditioned, which leads to numerical instability. To tackle such issue, some treatments, such as perturbing the diagonal elements of these matrices [30], will be incorporated. Since the objective function cannot be expressed in a closed form with respect to the hyper-parameters, gradient-based optimization algorithms may not perform well. Therefore, sampling-based stochastic optimization algorithms are preferred [24, 25, 31]. Specifically, here we adopt simulated annealing optimization (SAO) and particle swarm optimization (PSO) that are two representative algorithms.

Simulated annealing optimization (SAO) is a stochastic technique inspired by the physical annealing process, in which the metal cooling proceeds progressively until a lowest-energy state is reached [25]. At each virtual annealing temperature, the algorithm generates a new potential solution (or neighbor of the current state) to the problem considered by altering the current state, according to a predefined criterion. The acceptance of the new solution is then based upon the satisfaction of the Metropolis criterion [25]. Furthermore, an annealing schedule is selected to systematically decrease the temperature as the process continues. As the temperature decreases, the algorithm reduces the extent of its search to converge to a minimum or maximum. This approach was originally designed for solving single-objective optimization problem and was shown to be effective and convergent in various applications [32, 33].

Particle swarm optimization (PSO) is a population-based stochastic optimization algorithm motivated by intelligent cooperation of some animals referred to as swarm [24]. Each particle in the swarm keeps updating the search pattern in terms of learning experiences from its own and others until the solution is convergent. In particular, a collection of particles move iteratively throughout a pre-specified search region. At each iteration, the algorithm evaluates the objective function of each particle, and uses the objective information to decide the new velocity of each particle. The particles will move to the new locations with the velocities determined, and the algorithm then will reevaluate the objective function. Such procedure is repeated until all particles coalesce around one or several locations. Due to its intuitive background, implementation convenience, as well as the wide adaptability to different types of functions, its application to various domains has achieved great success [34, 35].

In this research, we integrate both SAO and PSO into MRGP training and compare their performance in terms of model updating accuracy and efficiency. SAO is carried out given the initial guess of hyper-parameters, while PSO is executed upon the defined search boundary of hyper-parameters. For the sake of comparison, we additionally apply a boundary constraint into the hyper-parameter search while using SAO. The related operating parameters are listed in Table 1.

### 3.3. Model updating practice

The frequency response differences between the finite element model simulation and the actual measurement are used as input information to update the model. Figure 3 shows an illustration of the FRF variations caused by the structural property change observed at one location/DOF. The differences at the 12 frequency points selected are employed for model updating. Without loss of generality, we also define the search space for each input parameters as $[-1, 1]$.

As indicated in Section 2.3, we establish an initial MRGP meta-model, and then resort to guided sampling for further finite element simulations which will then be used to update the MRGP meta-model. Moreover, we utilize all simulation results in the archived data to narrow down the input parametric space for error surface construction. In this case study, the MRGP meta-model will be able to emulate 72 (i.e., $6 \times 12$) response variables concurrently. As mentioned, we utilize SAO and PSO for training. To compare their performances directly, we employ the same training dataset at the initial stage. Without loss of generality, the error threshold $\varepsilon_s$ shown in Figure 1 is selected as 1% in this case study for convergence check. Table 2 lists the steps and the computational parameters employed.

Figure 4 shows the overall trend of error during iteration under two different optimization algorithms. Unless otherwise noted, the following results are split into two subsets corresponding to SAO and PSO, respectively. It is observed that overall errors can be minimized to 0.86% and 0.64% under SAO and PSO respectively after 9 iterations. The main computational cost of the updating process lies in the additional finite element simulations carried out. In this case study, owing to the adaptive sampling strategy, 9 iterations only require an additional 180 runs of the finite element simulation (according to Table 2), which exemplifies the high computational efficiency achieved. We can further look into the individual/local response error at each sensor location, as shown in Figure 5. Here for demonstration purpose, the individual/local response error is defined as the average error percentage of frequency response amplitudes under all excitation frequencies. For example, the local error of the *i*-th location/DOF is expressed as

$$\varepsilon_{e,i} = \frac{1}{p} \sum_{j=1}^{p} \left| \frac{\Delta u_{i,j}}{\bar{u}_{i,j}} \right| \tag{15}$$

Here $p = 12$, which represents the number of excitation frequency points. It can be seen that the average error trends under SAO and PAO exhibit similar patterns at all locations/DOFs, and will approach to nearly zero when the process converges. Clearly, the proposed framework can reach rapid error convergence.

The guided sampling of parameters to be updated in terms of MRGP-based error surface versus iteration is indicated in Figure 6. In the beginning, 20 initial samples are randomly generated from the full search space [-1,1]. Following the Step *d* in Table 2, at each iteration the best 20 model parameter samples will be determined according to the error surface feature. It is clearly observed that the variation of those samples will become smaller as iteration proceeds, which illustrates the adaptivity of this sampling technique. The trend of identified model parameters versus iteration hence can be obtained, and the parameter values at the last iteration is considered as the final solution (Figure 7). The horizontal dashed lines represent the actual values that can be used as a reference to examine the identification accuracy. The parameters 1, 2, 5 and 6 are identified quite accurately, while parameters 3 and 4 have certain discrepancies with respect to the actual values. From the optimization perspective, the complicated error surface with many local optima over different combinations of model parameters will render the solution search difficult to converge to global optima. From the physical perspective, model parameters 3 and 4 may not be sensitive to the selected frequency responses, which will be further discussed in the next subsection.

As emphasized, this proposed framework adopts an interactive, guided sampling strategy between MRGP meta-modeling and finite element simulation throughout the entire iterative process. The first guided sampling takes place when MRGP recommends the potentially best model parameter samples for finite element simulation according to the established error surface. The second one is to build a multivariate normal distribution based upon the archived data of finite element results, characterizing the parametric space for error surface prediction. This second aspect of guided sampling can be observed in Figure 8. The parametric space initially is specified as $[-1,1]$. Within this range, a uniform distribution is firstly adopted to generate the model parameter samples at the initial step. As iteration continues, the normal distributions will evolve with the updates of means and variances. As can be observed from Figure 8, most of distributions evolve properly, in which the updated mean values gradually approach to actual values and variances gradually reduce, pointing to a narrowed search space. The variances of distributions indicate not only the convergence speed, but also the confidence levels of the solution search. It can also be observed that the distribution mean of model parameter 3 at the final iteration still deviates from the actual value even though the convergence becomes slow. The distribution variances of parameter 3 and 4 are larger than those of others. The distribution evolutions to certain extent reflect the uneven identification errors (Figure 7).

The performances of the two optimization algorithms adopted can be compared through above results (Figures 4-8). Indeed, they perform quite similarly in terms of model updating accuracy. One important aspect to be examined is the computational efficiency. Figure 9 records the time costs for training MRGP meta-model at different iterations using SAO and PSO. It is worth nothing that the total number of iterations is one less than the numbers shown in the aforementioned discussions, since the analysis will terminate and skip the meta-model training once the error threshold is reached (Figure 1). The matrix operations, e.g., matrix inversion and multiplication etc., for objective evaluation, take the main computational time. It is hard to reduce such computational cost since the related matrices are not sparse.

As shown in Section 2, the sizes of matrices involved depend on the numbers of training data and response variables, and will increase as iteration proceeds. This leads to the increase of computational times per iteration (Figure 9). It is worth noting that optimization problem setup and parametric selection inevitably affect the computational times so it is impractical to facilitate a rigorous comparison.

Qualitatively, the computational costs of SAO and PSO are of the same order of magnitude, and PSO is slightly more efficient in MRGP hyper-parameter solution search than SAO.

*3.4. Investigation of performance robustness*

It is worth noting that randomness inevitably exists because of the stochastic nature of the algorithms involved in the proposed framework. Such randomness includes: 1) initial model parameter samples; 2) adaptive and guided sampling of model parameters at different iterations; and 3) MRGP meta-model training at different iterations. To examine the performance robustness, we carry out multiple runs of model updating and summarize the results statistically. Here we implement 5 runs and results are presented in Figures 10 and 11. Figure 10 shows the overall errors of 5 runs, which all are under the error threshold, i.e., 1%. The mean values of the error distributions are calculated as 0.83% and 0.81% using SAO and PSO, respectively. The identified model parameters are listed in Figure 11, which directly reflects the model updating accuracy. For all runs, the identification accuracy of model parameters 1, 2 and 6 are very good. Their error variations are generally small. While parameters 3, 4, and 5 are properly identified, there exist small errors. Overall, the results indicate that the proposed framework is robust in terms of model updating results. The errors also show consistency with those obtained drawn in Section 3.3 (Figure 7).

We also look into the computational cost under the aforementioned randomness. An important portion of computational cost is on finite element simulations to generate data under guided sampling. The case investigation is conducted on a desktop computer with Intel CPU E5-2640 @2.40GHz (2 processors). A single finite element simulation to evaluate FRF employed takes approximately 20 seconds. In all 5 runs, the results convergence is achieved with 9 to 12 iterations. Therefore, only a small number of finite element simulations, i.e., less than 250, are needed. This shows that the proposed method can indeed reach very good results efficiently.

To further demonstrate, we conduct an inverse identification of model parameters using a conventional approach to characterize the error surface over the entire parametric space. That is, we build a MRGP meta-model without adaptive and guided sampling. The solution can then be identified by sorting the errors of all model parameter samples. As model parameters are subject to uniform distribution within $[-1, 1]$, we generate 10,000 samples. 300 out of 10,000 model parameter samples are randomly selected and used in finite element simulations to compute the corresponding frequency response errors. These 300 data are used to train a MRGP meta-model which is then used to predict the error values over the original 10,000 samples. We use SAO and PAO, respectively, in MRGP meta-modeling, and execute 5 runs. Figure 12 shows the overall errors. Apparently, this conventional approach has much greater errors than those obtained from the proposed new method. SAO in this case performs slightly better than PSO. Figure 13 indicates that the model parameters identified have substantial discrepancy with respect to the actual values. Comparing these results with Figures 10 and 11, we can see that this conventional approach leads to greater variations of overall errors and parameter distributions. This is an indication that 300 data are insufficient to yield a credible MRGP meta-model for model updating. It is worth emphasizing that the reason we apply 300 data in the conventional method is that this already exceeds the computational time required for data preparation for the MRGP meta-modeling with guided sampling. In the latter, in all runs executed, only 180 to 240 finite element simulations (corresponding to the aforementioned 9 to 12 iterations) are needed to generate the data, and the results are much more accurate. For large-scale finite element model with higher dimension, the computational efficiency gain will be more significant.

### *3.5. Further discussion on model updating error*

As shown in Figures 7 and 11, the model parameter 1, 2, 5 and 6 are identified more accurately. To elucidate the underlying reason, we examine the correlation between the model parameters and frequency responses. We generate 10,000 model parameter samples, and compute the corresponding frequency responses through finite element analyses using the brute force Monte Carlo simulation. We adopt the so called Pearson correlation coefficient to quantify the correlation degree [36]. It essentially is a measure of the linear correlation between two variables defined as

$$\rho_{\boldsymbol{\alpha}_k, \mathbf{Z}_{i,j}} = \frac{\text{cov}(\boldsymbol{\alpha}_k, \mathbf{Z}_{i,j})}{\sigma_{\boldsymbol{\alpha}_k} \sigma_{\mathbf{Z}_{i,j}}} \tag{16}$$

where $\boldsymbol{\alpha}_k$ and $\mathbf{Z}_{i,j}$ represent, respectively, the data samples of the *k*-th model parameter and the frequency response amplitude at the *i*-th location under the *j*-th excitation frequency. cov(.) denotes the covariance function. $\rho_{\boldsymbol{\alpha}_k, \mathbf{Z}_{i,j}}$ generally falls into range [-1, 1]. The sign of correlation value defines the direction of the relation. The closer $\rho_{\boldsymbol{\alpha}_k, \mathbf{Z}_{i,j}}$ is to -1 (negative correlation) or 1 (positive correlation), the stronger the correlation between $\boldsymbol{\alpha}_k$ and $\mathbf{Z}_{i,j}$.

In this study, we have 6 model parameters and 12 frequency response amplitudes at each of 6 locations. Therefore, the Pearson coefficient values are evaluated 432 (i.e., $6 \times 12 \times 6$) times to capture all correlations. The details are shown in Figures 14-19. As can be observed, model parameters 1, 2 and 6 are highly correlated to the frequency responses measured at most locations. Model parameter 5 also has certain correlation with respect to the frequency responses in the vicinity of the second natural frequency measured at locations 5 and 6. This is the reason that these model parameters can be accurately identified/updated under the current problem setup. In comparison, the influence of model parameters 3 and 4 on frequency responses acquired are insignificant. The location of response measurement certainly affects the identification accuracy. This analysis reveals the underlying physical reason for the errors in model updating practice. In actual implementations, such analysis may provide insight to placing sensors at proper locations.

## 4. Concluding Remarks

A new finite element model updating framework built upon the multi-response Gaussian process (MRGP) meta-model is developed in this research. MRGP meta-model is utilized to efficiently characterize the frequency response errors between finite element simulation results and actual measurement. In particular, the MRGP meta-model is continuously updated by introducing additional sampled model parameters and related frequency response errors computed from finite element model. This framework enables an interactive, guided sampling strategy between MRGP meta-modeling and finite element simulation, which expedites the search convergence of unknown model parameters. Two stochastic optimization techniques, i.e., simulated annealing and particle swarm are adopted and their performances are compared. Comprehensive case studies are carried out. The results demonstrate that the proposed framework can yield very accurate model updating results with improved efficiency, and the performance is robust.

**Acknowledgment**

This research is supported by National Science Foundation under grant CMMI-1825324.

**Table 1. Computational parameters of optimization algorithms for MRGP meta-model training**

| Simulated annealing optimization (SAO) | | Particle swarm optimization (PSO) | |
|---|---|---|---|
| Upper bound of $l$ | 0 | Upper bound of $l$ | 0 |
| Lower bound of $l$ | $\tilde{l}_{SAO,i}$ | Lower bound of $l$ | $\tilde{l}_{PSO,i}$ |
| Upper bound of $\sigma_f$ | 0 | Upper bound of $\sigma_f$ | 0 |
| Lower bound of $\sigma_f$ | 10 | Lower bound of $\sigma_f$ | 10 |
| Initial value of $l$ | 1 | Swarm size | 100 |
| Initial value of $\sigma_f$ | 1 | Maximum evaluation number | 10,000 |
| Initial temperature | 100 | | |
| Target temperature | 3 | | |
| Temperature descent slope | 0.8 | | |
| Iteration number at each temperature | 400 | | |

Note:
1. Hyper-parameter $l$ (i.e., lengthscale) is a positive value according to the nature of squared exponential covariance function. The lower bound is always set as 0 regardless of iteration. The upper bounds of $\tilde{l}_{SAO,i}$ and $\tilde{l}_{PSO,i}$ in the $i$-th iteration are determined by the optimized hyper parameters in previous iteration.

2. Hyper-parameter $\sigma_f^2$ (i.e., signal variance) usually has small search space because large value of $\sigma_f^2$ is not helpful for data fitting, and meanwhile will cause the numerical instability. We thus set a fixed bound for $\sigma_f^2$.

**Table 2. Implementation procedure**

| |
|---|
| *a*. Randomly generate $q$=20 parameter samples via uniform distribution within full search space [-1,1] |
| *b*. Conduct finite element analysis to produce corresponding frequency response data; calculate the frequency response discrepancies/errors between sampled finite element simulation results and actual response measurement. One model parameter sample together with its errors are employed as one training data point. |
| *c*. Use $q$=20 training data to establish MRGP meta-model, and predict the frequency response errors over $w$=10,000 samples parameterized from uniform distribution with full search space [-1,1] |
| *d*. Sort the parameter samples in terms of the defined overall error (Equation (6b)), **and select top $s$=20 samples with smaller overall errors**. |
| *e*. Repeat *b* and evaluate the overall error. If the smallest overall error is smaller than the threshold defined, stop; otherwise execute *f*. |
| *f*. At this point, total $q+s$=40 samples are evaluated through finite element analysis; select top $z$=20 samples out of $q+s$=40 with smaller overall errors and calculate their means and variances. |
| *g*. Use above $q+s$=40 training data to establish MRGP meta-model, and predict the frequency response errors under $w$=10,000 new samples **parameterized from a multivariate normal distribution upon the mean and variance** obtained in *f*. |
| *h*. repeat *d*. |

Note:

1. The numbers in steps *f* and *g* will increase during iteration. For example, for the *i*-th iteration, the numbers in steps *f* and *g* respectively are $20 \times (i-1) + 20$ and $20 \times (i-1)$ (the numbers shown above are related to the second iteration).

2. Bolded texts indicate those related to adaptive sampling.

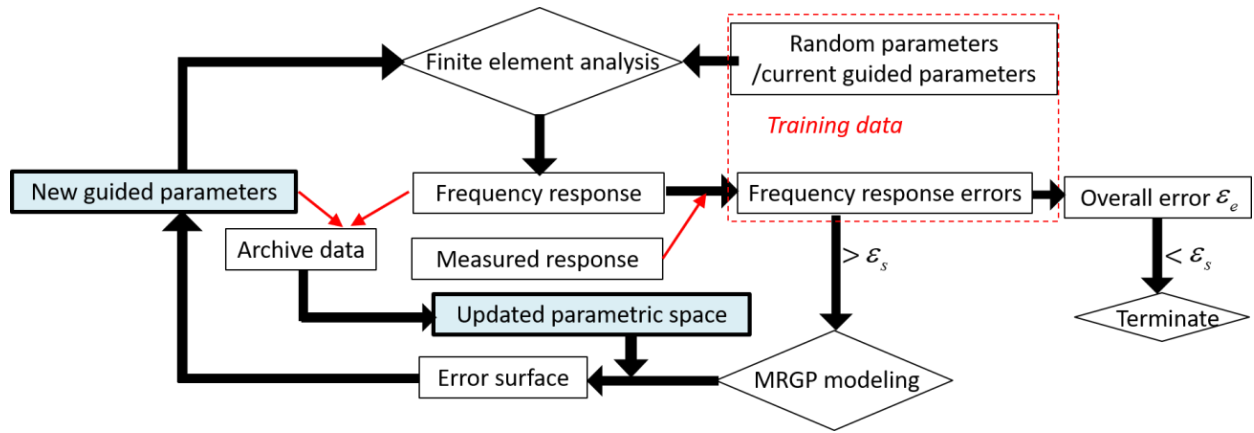

Figure 1. Computational procedure of the model updating framework.

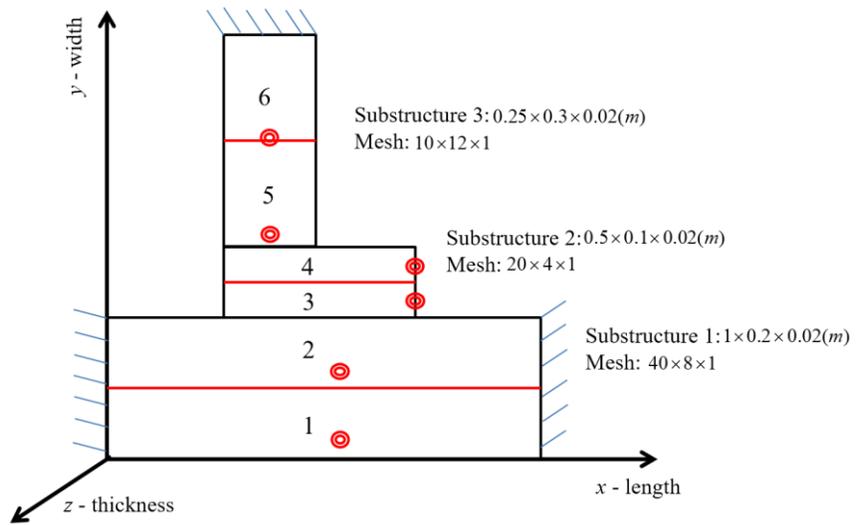

Figure 2. Benchmak structure investigated and mesh/segment set-up.

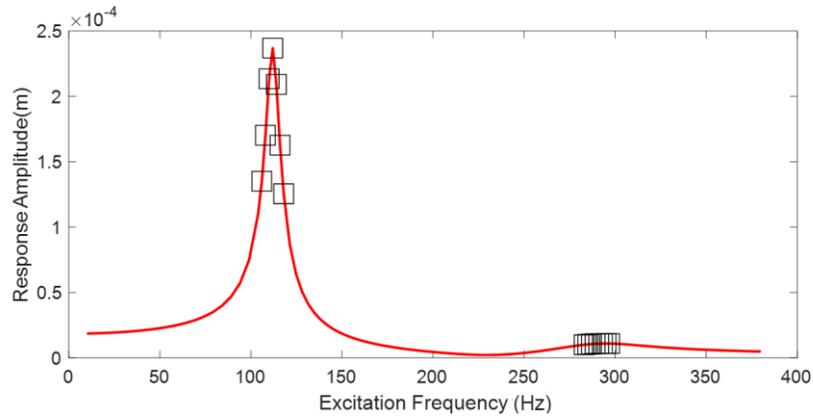
(a)

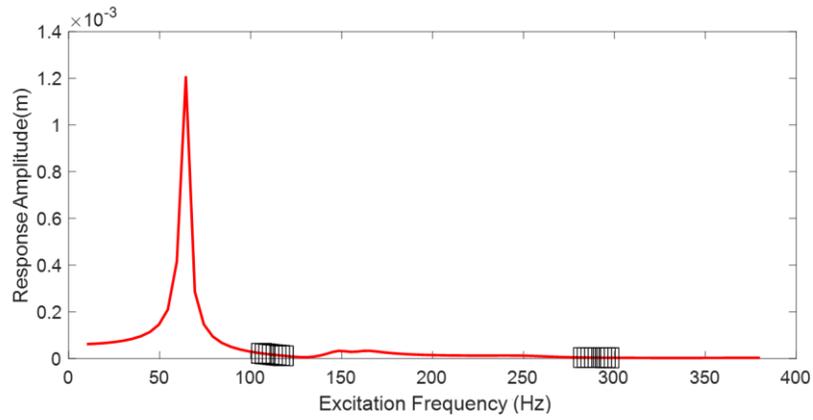
(b)

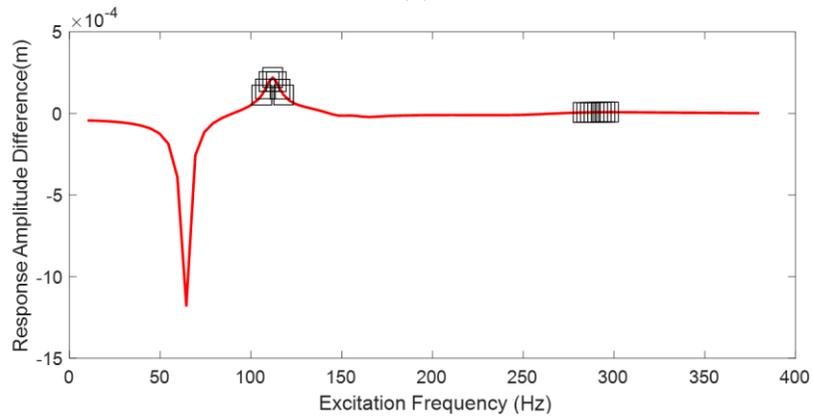
(c)

Figure 3. Illustration of FRF difference at location/DOF 5 (a) FRF of actual structure with model parameters: $\boldsymbol{\gamma} = [-0.6, -0.1, -0.3, -0.2, -0.2, -0.4]$; (b) FRF of structure with model parameters: $\boldsymbol{\gamma} = [-0.8, -0.8, -0.8, -0.8, -0.8, -0.8]$ (c) FRF difference. Squares denote the selected response points for model updating.

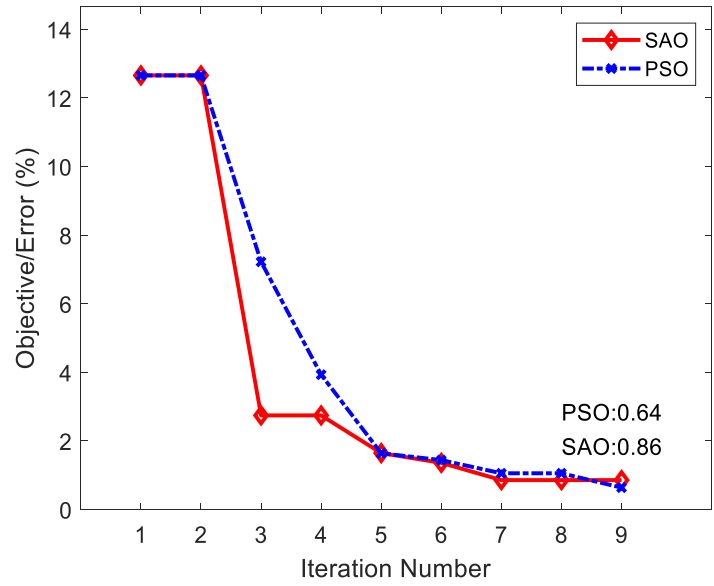

Figure 4. Error trend during iteration.

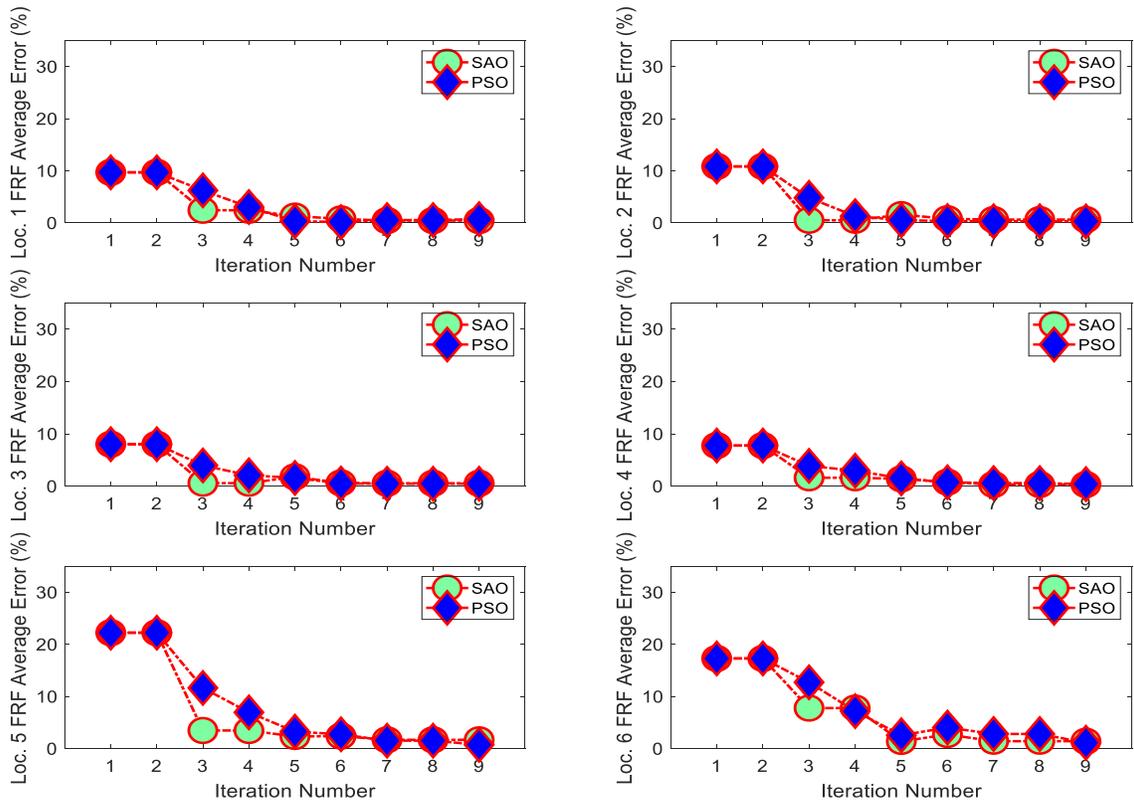

Figure 5. FRF average error trends during iteration.

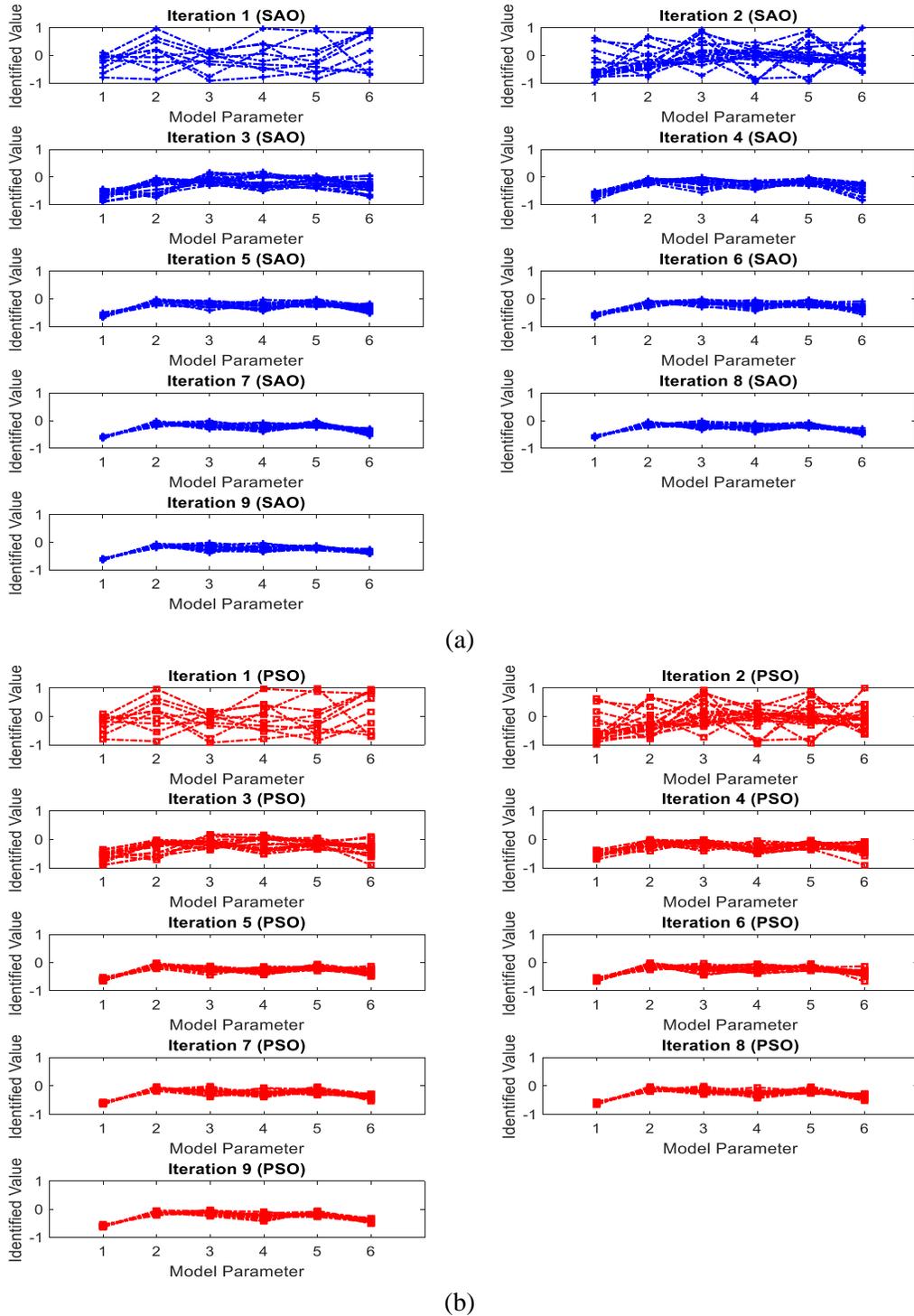

Figure 6. Guided model parameter sampling from MRGP-based error surface versus iteration (a) Simulated annealing optimization (SAO); (b) Particle swarm optimization (PSO). Initial random search starts within the full parametric space [-1, 1] characterized by a uniform distribution.

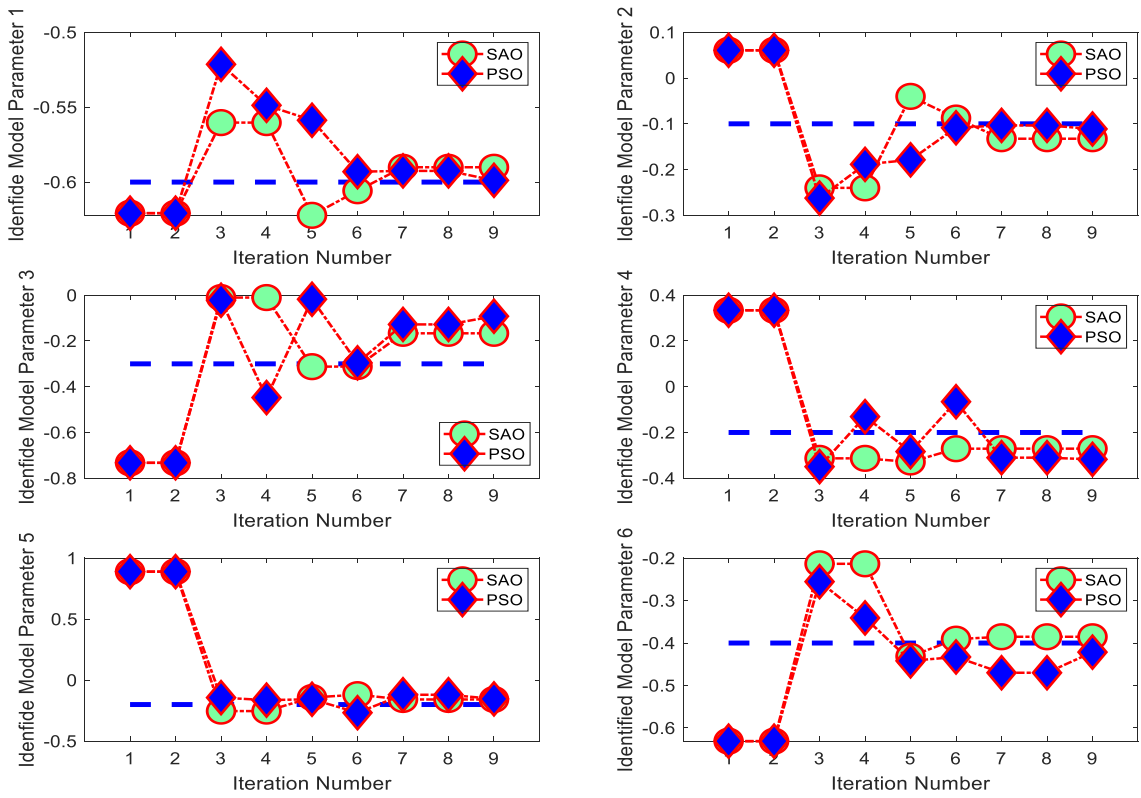

Figure 7. Updated model parameters during iteration. Dashed lines represent the actual values.

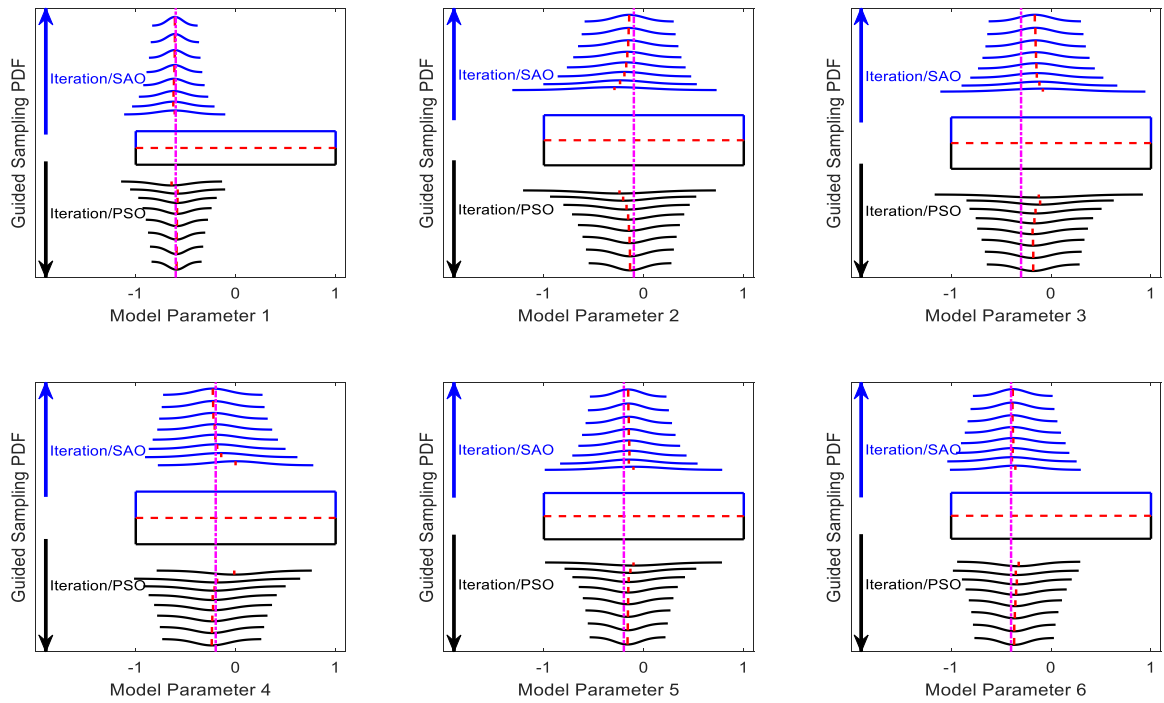

Figure 8. Evolutions of guided normal distributions versus iteration. Vertical dashed lines represent the actual model parameters; horizontal dashed lines split the results of SAO and PSO.

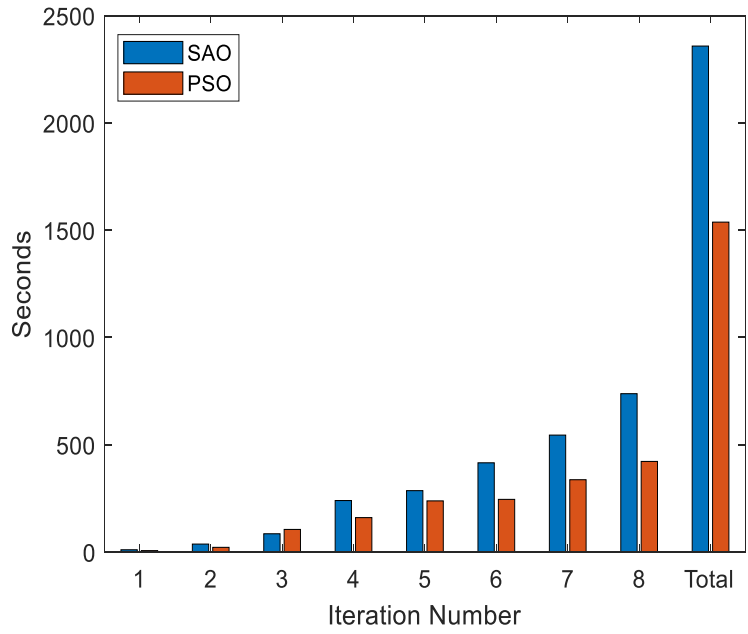
Figure 9. Computational cost of MRGP meta-model training.

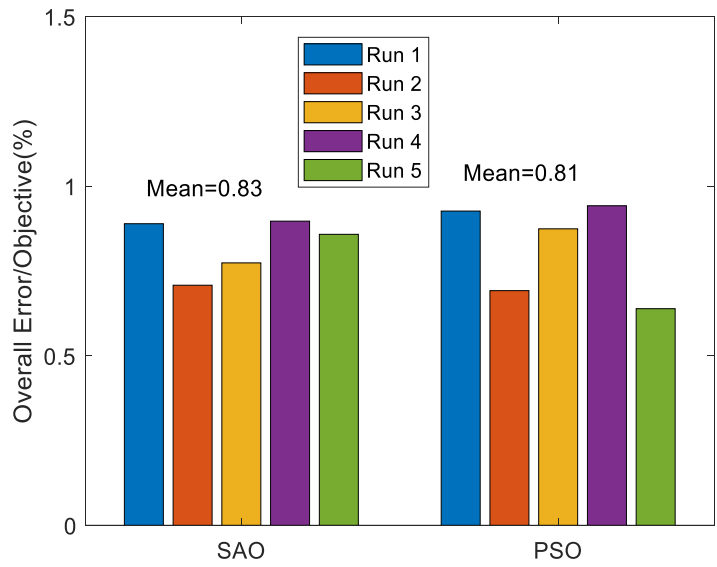
Figure 10. Statistics of overall errors under different runs.

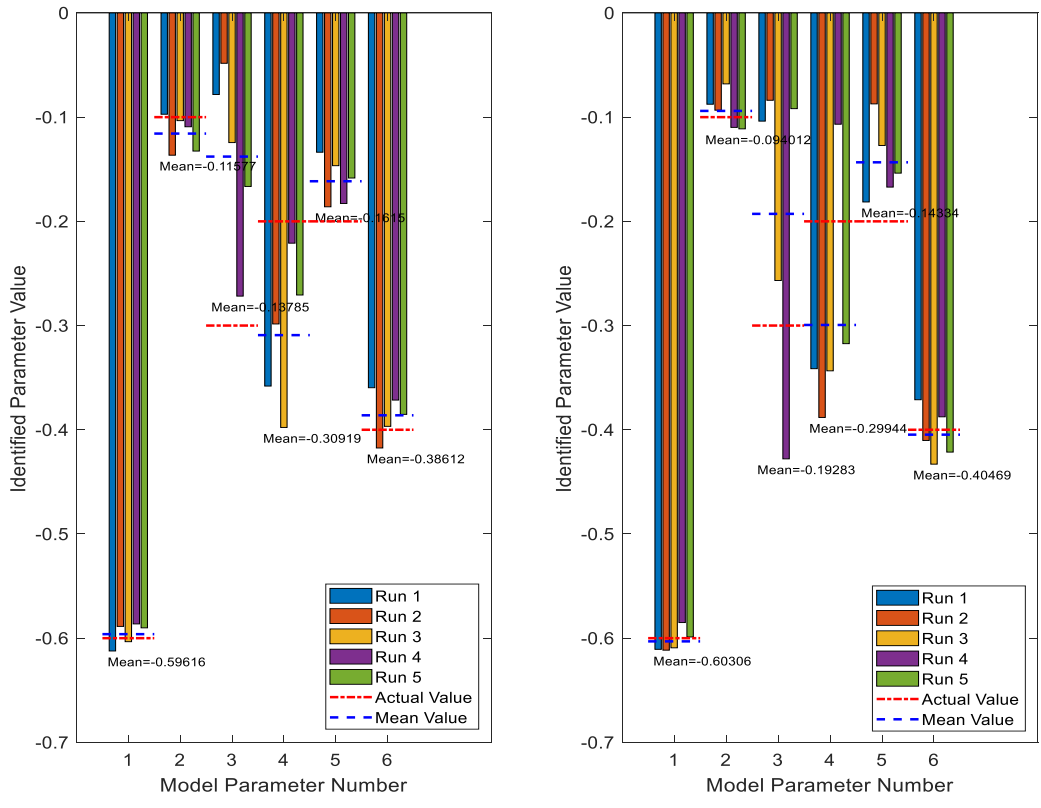

Figure 11. Statistics of model parameters identified under different runs.

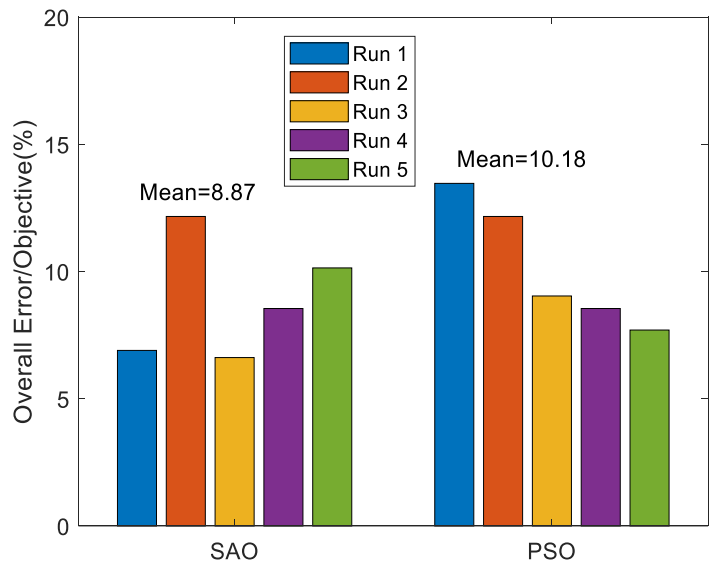

Figure 12. Statistics of overall errors under different runs using conventional approach.

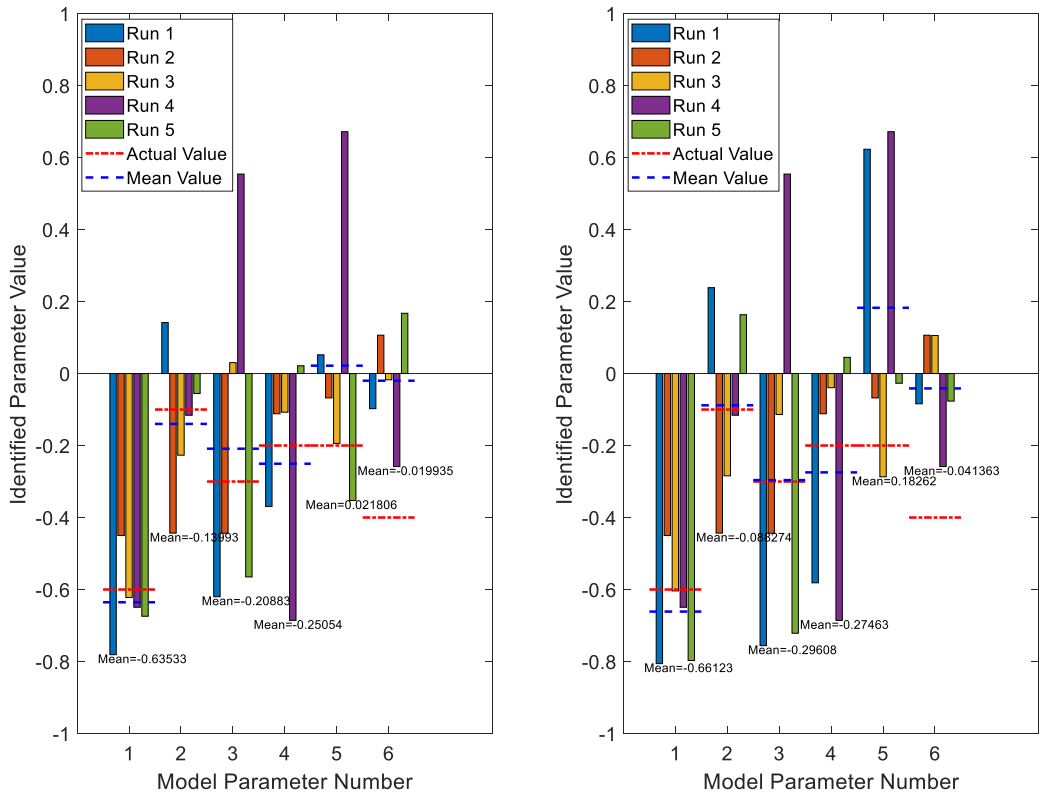

Figure 13. Statistics of model parameters identified under different runs using conventional approach.

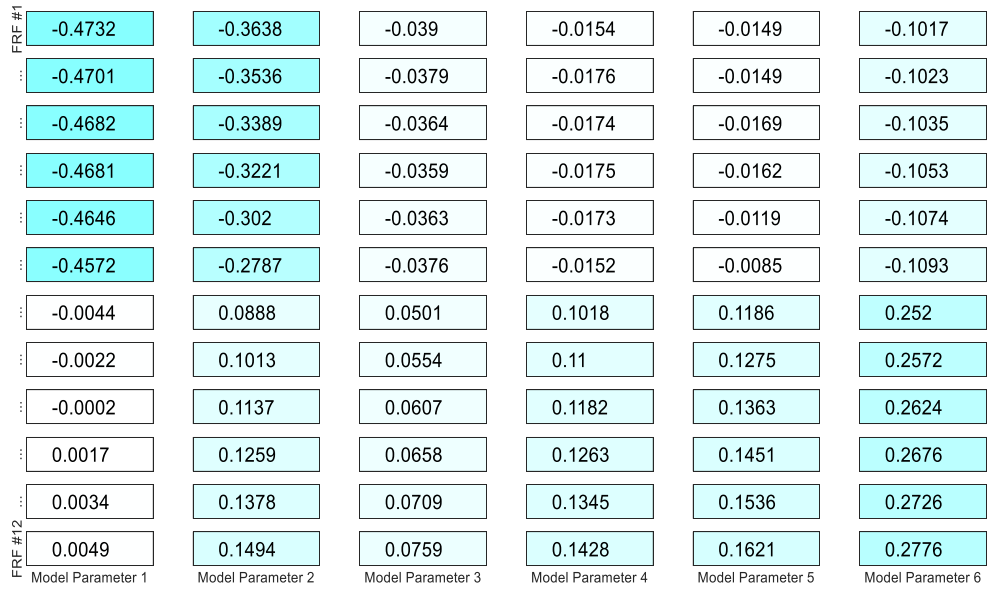

Figure 14. Correlation analysis: model parameters versus frequency response amplitudes at location 1.

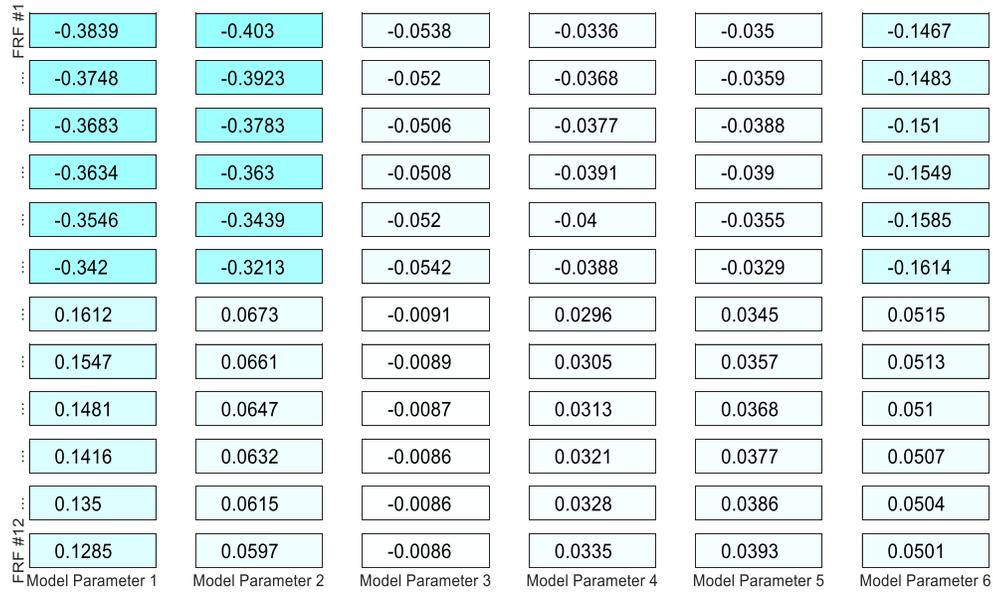

Figure 15. Correlation analysis: model parameters versus frequency response amplitudes at location 2.

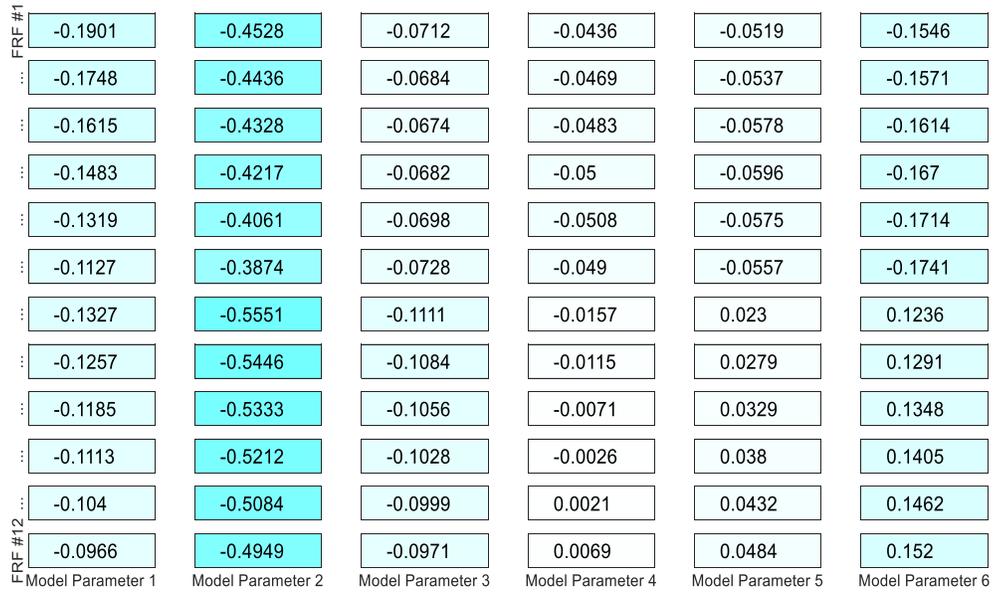

Figure 16. Correlation analysis: model parameters versus frequency response amplitudes at location 3.

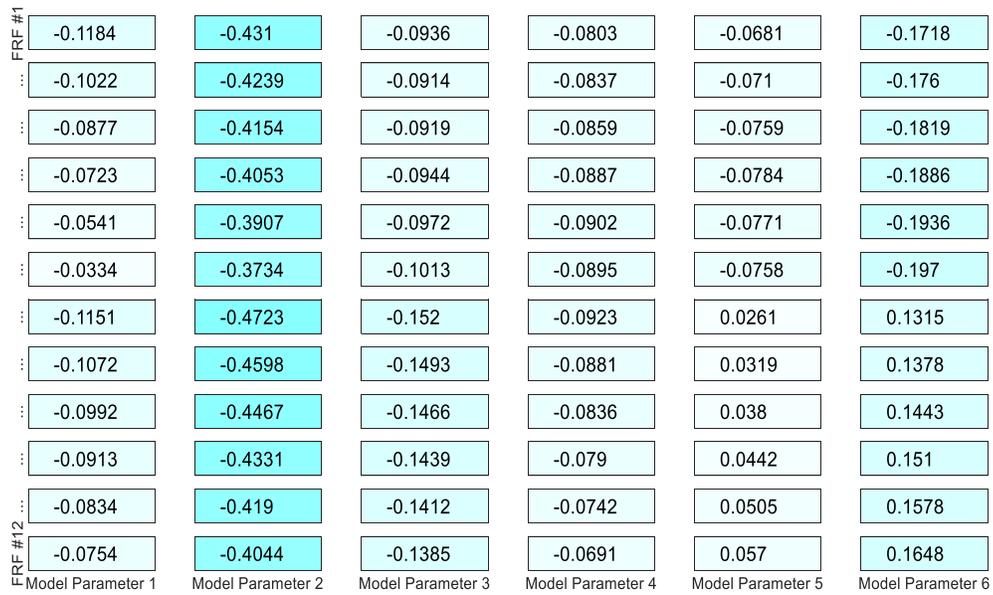

Figure 17. Correlation analysis: modelparameters versus frequency response amplitudes at location 4.

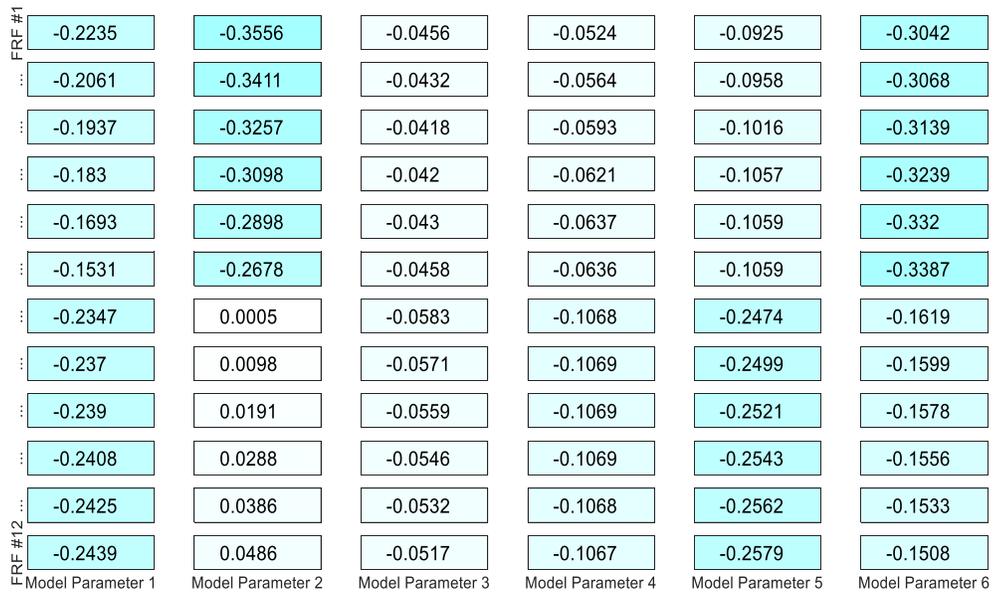
Figure 18. Correlation analysis: model parameters versus frequency response amplitudes at location 5.

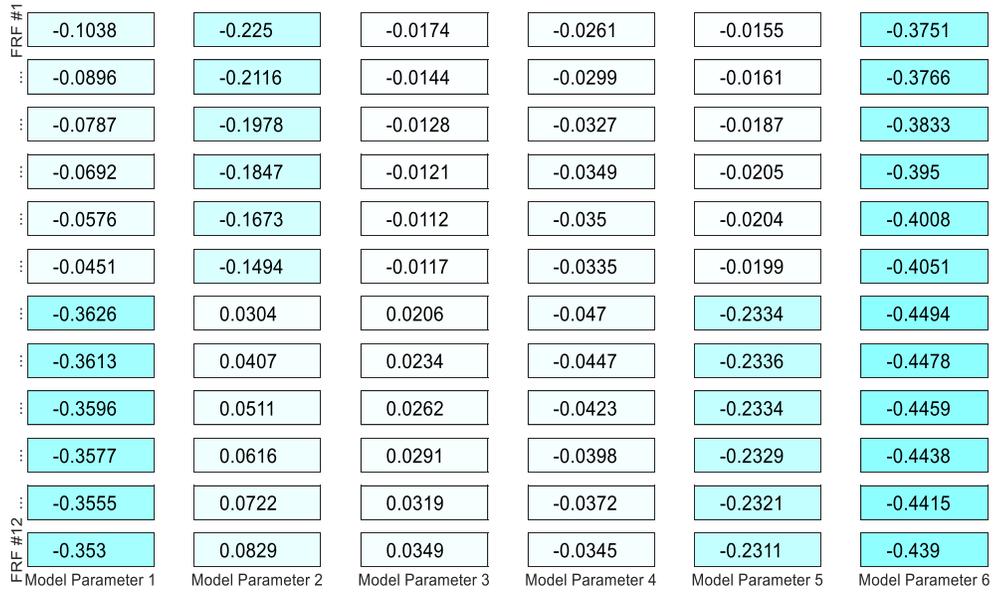

Figure 19. Correlation analysis: model parameters versus frequency response amplitudes at location 6.